\documentclass[prx,reprint,superscriptaddress,twocolumn,showkeys]{revtex4-1}
\usepackage[utf8]{inputenc} 
\usepackage{amsmath}
\usepackage{braket}
\usepackage{qcircuit}
\usepackage{graphicx}
\usepackage{pgfplots}
\pgfplotsset{compat=1.15}
\usepackage{csquotes}
\usepackage{hhline}
\usepackage{amssymb}

\usepackage{dcolumn}
\usepackage{tabularx}
\usepackage[colorlinks=true,linkcolor=blue,citecolor=blue,urlcolor=blue]{hyperref}
\usepackage{braket}
\usepackage{float}
\usepackage{enumerate}
\usepackage{listings}
\usepackage{color}
\usepackage{subfigure}
\usepackage{placeins}
\usepackage{natbib}
\definecolor{mygreen}{rgb}{0,0.6,0}
\definecolor{mygray}{rgb}{0.5,0.5,0.5}
\definecolor{mymauve}{rgb}{0.58,0,0.82}

\newcolumntype{C}{>{\centering\arraybackslash}X}

\begin{document}
\title{Experimental realization of quantum teleportation of arbitrary single and two-qubit states via hypergraph states}
\author{Atmadev Rai}
\email{atmadev85213@gmail.com}
\affiliation{Department of Physics,\\ Indian Institute of Technology (ISM) Dhanbad India}
\author{Bikash K. Behera}
\email{bikas.riki@gmail.com}
\affiliation{Bikash's Quantum (OPC) Pvt. Ltd., Balindi, Mohanpur 741246, West Bengal, India}

\begin{abstract}
Here we demonstrate quantum teleportation through hypergraph states, which are the generalization of graph states and due to their non-local entanglement properties, it allows us to perform quantum teleportation. Here we design some hypergraph states useful for quantum teleportation and process the schemes for quantum teleportation of single-qubit and two-qubit arbitrary states via three-uniform three-qubit and four-qubit hypergraph states respectively. We explicate the experimental realization of quantum teleportation of both single and two-qubit arbitrary states. Then we run our quantum circuits on the IBM quantum experience platform, where we present the results obtained by both the simulator and real devices such as ``ibmq\_qasm\_simulator" and ``ibmq\_16\_melbourne" and calculate the fidelity. We observe that the real device has some errors in comparison to the simulator, these errors are due to the decoherence effect in the quantum channel and gate errors. We then illustrate the experimental and theoretical density matrices of a teleported single and two-qubit states.
\end{abstract}

\maketitle

\section{Introduction}
One of the most curious characteristics of quantum mechanics that always has been at the forefront of all the discussions is entanglement \cite{nielsen2002quantum}. Quantum entanglement is one of the fundamental resources of quantum information theory that plays an important role in quantum information science and quantum computing. It is the key element of most of the applications of quantum information and computation, such as quantum teleportation \cite{Murli2008,Murli2009}, quantum secret sharing \cite{HMBB1999,GOTD2000}, quantum cryptography \cite{BCHB1984}, quantum error correction \cite{Walker2008}, quantum dense coding \cite{BCHW1992} and many more \cite{Moul2016,Behera2017,Zhang2006}.

Quantum state tomography \cite{JDFV2001} and quantum process tomography \cite{DWCJ2001} are the two elementary methods that are significantly important for quantum computation and information. Quantum state tomography is a process to investigate the state of a system by overcoming the hidden nature of quantum mechanics since we can not determine a state measuring it directly. Quantum state tomography performs repeated preparations of the same quantum state and measures on a different basis to obtain a complete state description. We have also used quantum state tomography to determine the experimental density matrix.

There is open access to the prototypes of quantum computers provided by IBM quantum experience for testing and simulating the quantum algorithms; it allows the users to perform quantum computing applications and design quantum circuits. IBM QE is widely used in many real experiments such as quantum simulation \cite{GDAP2018,VPKJ2018,SMFP2017,WBJR2018}, quantum cryptography \cite{BBPK2017,PMMTL2018,MMSK2017}, quantum error correction \cite{GDAP2018,RJHD2018}, quantum algorithms \cite{GMDS2018,JHAD2018,SMST2017,GSMB2018} etc. Cluster states \cite{NMA2006,DPXY2006} are commonly used for quantum teleportation; these are highly entangled state and are considered as a particular case of graph states. Cluster states have a great significance for quantum teleportation and one-way quantum computing \cite{WALTP2005,RRBH2001}.

One of the fundamental phenomena of quantum information is quantum teleportation utilizing the entanglement resources; it provides a key primitive across many quantum information tasks and plays an important role in building blocks for quantum technology. It has a big hand behind the progress of quantum communication, quantum computing and many quantum technologies. It allows the transfer of quantum information into an otherwise unreachable space. It is also in high demand for the quantum storage of highly secret data, such as DNA data, we can securely transfer the data by photons into quantum memories. Quantum teleportation is a technique for moving quantum states anywhere, even if there is no quantum communication channel linking both sender and receiver ends. It was firstly introduced in 1993 by Bennett \cite{BBCJ1993}, and later it was demonstrated experimentally by Bouwmeester \emph{et al.} \cite{BDPM1997} and Boschi \emph{et al.} \cite{BDBS1998}.

Our paper is structured as follows. In section-\ref{RSP_Sec2}, we briefly discuss hypergraph states. Section-\ref{RSP_Sec3} describes the schemes for teleportation of arbitrary single and two-qubit states; in section-\ref{RSP_Sec4}, we talk about IBM quantum experience (IBM QE), the platform where we can design and run our quantum circuits. Section-\ref{RSP_Sec5} deals with quantum state tomography of our experimental data. The results obtained from our experiments is shown in section-\ref{RSP_Sec6}. We conclude in section-\ref{RSP_Sec7} by providing future directions for this work.

\section{Hypergraph state\label{RSP_Sec2}}
Quantum hypergraph states are first discussed in ref. \cite{RMNP2013}. These are a group of highly entangled quantum states. Recently, hypergraph states are drawing much attention; the local unitary symmetries of hypergraphs and their local Pauli stabilizers are discussed in ref. \cite{LYOM2017}. The relationship between LME states and hypergraphs under local unitary transformations is given in ref. \cite{lyons2015local}. Multipartite entanglement in hypergraphs is discussed in ref. \cite{QURI2013}.

The mathematical theory of the graph is considered as the basis with which to introduce graph states. The concept of graph states is further generalized to hypergraph states. Hypergraph states are considered as a generalization of graph states \cite{Zwang2013}, whereas only two qubits in a graph state are involved in one interaction, an edge in the corresponding graph can only join two vertices. In contrast, we can include any number of qubits for interaction in a hypergraph state. Thus, a hypergraph can have edges that join more than one vertex, the edges that join more than two vertices are known as `hyperedges'. The cardinality of a hyperedge is the number of vertices within this edge, denoted by k. A hyperedge of cardinality k is referred to as a k-hyperedge. A hypergraph is referred to as constant cardinality k if all of the hyperedges in the hypergraph are k-hyperedges. Thus such a hypergraph can also be called a k-uniform hypergraph \cite{Guhne2014}. A hypergraph state is a set $H_n$=\{V, E\} of n vertices V, and hyperedges E of any order k, i.e., a hyperedge can have k number of vertices. A quantum hypergraph is created from a mathematical hypergraph by first assigning a qubit to each vertex and initializing them by applying the Hadamard gate on each qubit. If $j_1,j_2,j_3,...j_k$ are connected by a k hyperedge, then we perform $C^kZ_{j_1,j_2,j_3,...j_k}$ over the initialized n-qubit state $\ket{+}^{\otimes n}$ to get the hypergraph state given by,

\begin{equation}
    \ket{H}_n=\prod_{k=1}^{n}\prod_{j_1,j_2,j_3,...j_k}^{E}C^{k}Z_{j_1,j_2,j_3,...j_k}\ket{+}^{\otimes n}\nonumber\\
\end{equation}

Whereas the k-uniform hypergraph state $H_k=\{V, E\}$ is a set of vertices V and hyperedges E, where each edge connects with exactly k number of vertices, called k-uniform hypergraph state. A simple graph state is a 2-uniform hypergraph state because the edge of a graph only connects two vertices. We can obtain a k-uniform hypergraph state by following the same procedure as discussed for generalized hypergraph states by assigning a qubit to each vertex and initializing it to $\ket{+}$ state. We perform the controlled-Z operation to each k-hyperedges. If $j_1,j_2,j_3,...j_k$ are connected by k-hyperedge, then we perform $C^kZ_{j_1,j_2,j_3,...j_k}$ over the initialized n-qubit state $\ket{+}^{\otimes n}$ to get the hypergraph state given by,

\begin{equation}
    \ket{H}_k=\prod_{\{j_1,j_2,j_3,...j_k\}\in E}C^{k}Z_{j_1,j_2,j_3,...j_k}\ket{+}^{\otimes n}\nonumber\\
\end{equation}

\section{Scheme for quantum teleportation \label{RSP_Sec3}}
\subsection{Quantum teleportation of an arbitrary single-qubit state using a three-qubit hypergraph state}
Here we have considered a three-qubit three-uniform hypergraph state which we use as a quantum channel to teleport an arbitrary single-qubit state. A three-qubit three-uniform hypergraph state is given by,

\begin{eqnarray}
    \ket{h}=\frac{1}{2\sqrt{2}} (\ket{000}+\Ket{001}+\Ket{010}+\ket{011}\nonumber\\
    +\ket{100}+\ket{101}+\ket{110}-\ket{111})
\label{RSP_Eq1}    
\end{eqnarray}

\begin{figure}
\Large\Qcircuit @C=2em @R=1.3em {
&\lstick{\ket{0}} &\gate{H} & \ctrl{1}& \ctrl{1} & \qw  \\
&\lstick{\ket{0}} &\gate{H}& \ctrl{1}& \ctrlo{1}& \qw   \\
& \lstick{\ket{0}} &\gate{H} & \gate{Z} & \gate{Z}&\gate{H} 
}
\caption{Quantum circuit generating three-qubit uniform hypergraph state $\ket{H}_{a_1,a_2,b}$ after applying Hadamard gate on the third qubit.}
\label{Fig1}
\end{figure}
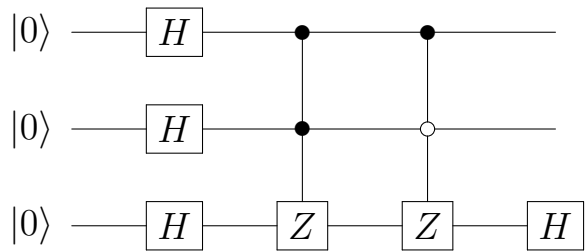

After reducing this state as shown in Fig. \ref{Fig1} to get

\begin{eqnarray}
    \ket{H}_{a_1a_2b}&=&\frac{1}{2}(\ket{000}+\ket{010}+\ket{101}+\ket{111})\nonumber\\
    \label{RSP_Eq2}
\end{eqnarray}

The first and second qubits belong to Alice, denoted as subscript $a_1$ and $a_2$, whereas the third qubit belongs to Bob, denoted as $b$.

\begin{figure*}
\centering
  \Qcircuit @C=2.4em @R=2.5em {
&\lstick{\ket{0}} &\gate{U3}&\qw& \qw &\qw&\barrier{3}\qw& \ctrl{1}& \ctrl{2} & \gate{H} &\qw&\qw&\qw&\ctrl{3}&\qw &\qw\\
&\lstick{\ket{0}} &\gate{H} & \ctrl{1}& \ctrl{1}&\qw&\qw  & \targ  &\qw  &\qw&\qw&\ctrl{2}&\qw&\qw&\qw&\qw \\
&\lstick{\ket{0}} &\gate{H}& \ctrl{1}& \ctrlo{1} &\qw&\qw&\qw &\targ&\qw&\gate{H}&\qw&\qw&\qw &\qw&\qw \\
& \lstick{\ket{0}} &\gate{H} & \gate{Z} & \gate{Z} &\gate{H} &\qw&\qw &\qw &\qw&\qw&\targ&\qw&\gate{Z}&\meter&\qw
}
    \caption{A generalised circuit for quantum teleportation of single qubit using three-qubit uniform hypergraph state. }
    \label{Fig2}
\end{figure*}
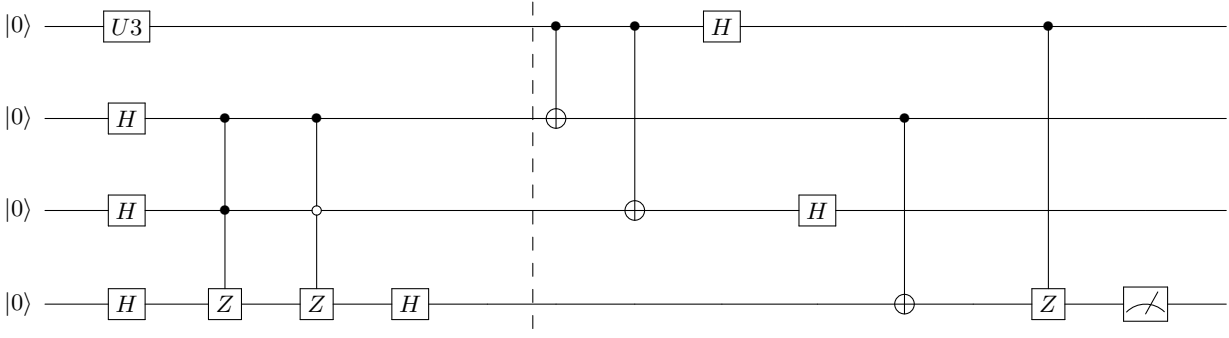

In this scheme, Alice wants to teleport any single-qubit state $\ket{\Psi}_a=(\alpha\ket{0}+\beta\ket{1})$ through the three-qubit hypergraph state, where $|\alpha|^2+|\beta|^2=1$. The joint state of the arbitrary single-qubit state and the three-qubit hypergraph state can be written as $\ket{\Psi}_{aa_1a_2b}=\ket{\Psi}_a\otimes\ket{H}_{a_1a_2b}$. Let us briefly discuss this teleportation process; Alice has a qubit of an arbitrary state which she wants to teleport and two qubits of the entangled hypergraph state as mentioned earlier. First, she performs controlled-Not (CNOT) gate on qubits ($a, a_1$) and qubits ($a, a_2$), where the qubit a works as controlling qubit and qubits $a_1$ and $a_2$ works as target qubits. Then she performs Hadamard gate on her qubit $a$; Alice measures her qubit on a computational basis and classically informs Bob about her measurement outcome. Now, Bob can obtain the unknown quantum state by performing some unitary transformations on his part of qubit b. In Fig. \ref{Fig2}, we can see the generalized circuit for teleportation of arbitrary single-qubit state by using a three-qubit hypergraph state.

\textbf{Circuit decomposition}: Let, single qubit arbitrary state which Alice wants to teleport is given by, $\Ket{\Psi}_a=(\alpha\Ket{0}+\beta\ket{1}$) and the uniform three-qubit hypergraph state which Alice and Bob share is given as $\Ket{H}_{a_1a_2b}$ in Eq. \ref{RSP_Eq2}. Now, Alice combines a single qubit arbitrary state to her part of an entangled qubit from hypergraph state. First of all, Alice performs CNOT gate on qubits ($a, a_1$) and qubits ($a, a_2$), so the joint state $\ket{\Psi}_{aa_1a_2b}$ is changed to,

\begin{eqnarray}
\ket{\Psi'}_{aa_1a_2b}=\frac{1}{2}(\alpha\ket{0}\otimes(\ket{000}+\ket{010}+\ket{101}+\ket{111})\nonumber\\
+\beta\ket{1}\otimes(\ket{110}+\ket{100}+\ket{011}+\ket{001}))\nonumber \\ \label{RSP_Eq3}
\end{eqnarray}

Then Alice applies a Hadamard gate on her qubit $a$ so the new state becomes,

\begin{eqnarray}
\ket{\Psi''}_{aa_1a_2b}=\frac{1}{2\sqrt{2}}(\alpha(\ket{0}+\ket{1})\otimes(\ket{000}+\ket{010}+\ket{101}\nonumber\\+\ket{111})\nonumber\\
+\beta(\ket{0}-\ket{1})\otimes(\ket{110}+\ket{100}+\ket{011}+\ket{001})\nonumber \\ \label{RSP_Eq4}
\end{eqnarray}

After expanding the above equation and rearranging, Alice performs a Hadamard operation on her second qubit of the hypergraph state, i.e., $a_2$, to collapse her state to get,

\begin{eqnarray}
\Ket{\Psi'''}=\frac{1}{2\sqrt{2}}(\ket{000}_{aa_1a_2}\otimes\ket{\Psi^1}_b+\ket{100}_{aa_1a_2}\otimes\ket{\Psi^2}_b\nonumber\\
+\ket{010}_{aa_1a_2}\otimes\ket{\Psi^3}_b+\ket{110}_{aa_1a_2}\otimes\ket{\Psi^4}_b)\nonumber\\
\label{RSP_Eq5}
\end{eqnarray}
where,
\begin{eqnarray}
\ket{\Psi^1}=(\alpha\ket{0}+\beta\ket{1})_b\nonumber\\
\ket{\Psi^2}=(\alpha\ket{0}-\beta\ket{1})_b\nonumber\\
\ket{\Psi^3}=(\alpha\ket{1}+\beta\ket{0})_b\nonumber\\
\ket{\Psi^4}=(\alpha\ket{1}-\beta\ket{0})_b\nonumber\\
\label{RSP_Eq6}
\end{eqnarray}

Now Alice measures her qubit on a computational basis and sends the classical information to Bob. Depending upon Alice’s classical information, Bob performs corresponding unitary operations as shown in Table \ref{TableI} to recover the arbitrary single-qubit state in its qubit.

\begin{table}
\begin{tabular}{|c|c|c|} 
\hline
Alice's outcome & Bob obtains & Bob's Operations\\
\hline
$\Ket{000}$ & $(\alpha\ket{0}+\beta\ket{1})$ & $I$\\
$\Ket{100}$ & $(\alpha\ket{0}+\beta\ket{1})$ & $Z$\\
$\Ket{010}$ & $(\alpha\ket{0}+\beta\ket{1})$ & $X$\\
$\Ket{110}$ & $(\alpha\ket{0}+\beta\ket{1})$ & $ZX$\\
\hline
\end{tabular}
\caption{Bob’s unitary operations corresponding to Alice's measurement results}
\label{TableI}
\end{table}
For example, if Alice measures her qubit and it turns out to be $\Ket{010}$, she then sends classical information to Bob. When Bob knows that Alice’s qubit after the measurement is $\Ket{010}$ then he applies an X gate on his qubit to obtain the actual state sent by Alice. So, this protocol for teleportation is deterministic, i.e., the probability of success is 100\%.
\subsection{Quantum teleportation of an arbitrary two-qubit state using a three-uniform four-qubit hypergraph state}
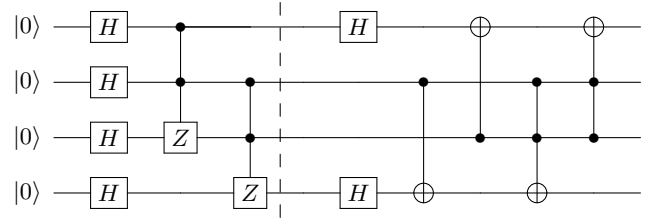
\begin{figure}[H]
\centering
\Qcircuit @C=1.5em @R=1em {
&\lstick{\ket{0}} &\gate{H} & \ctrl{1}&\qw\barrier{3}\qw&\qw&\gate{H}&\qw&\targ&\qw&\targ&\qw \\
&\lstick{\ket{0}} &\gate{H}&\ctrl{1}&\ctrl{1}&\qw&\qw&\ctrl{2}&\qw&\ctrl{1}&\ctrl{-1}&\qw\\
& \lstick{\ket{0}} &\gate{H} &\gate{Z}&\ctrl{1} &\qw &\qw&\qw&\ctrl{-2}&\ctrl{1}&\ctrl{-1}&\qw\\
&\lstick{\ket{0}} &\gate{H} &\qw & \gate{Z} &\qw&\gate{H}&\targ&\qw&\targ&\qw&\qw
}
    \caption{Quantum circuit generating three-uniform four-qubit hypergraph state $\ket{H_4}_{a_1a_2b_1b_2}$ after applying two Hadamard gate on a first and fourth qubit and two CNOT and two CCNOT operations as shown.}
    \label{Fig3}
\end{figure}
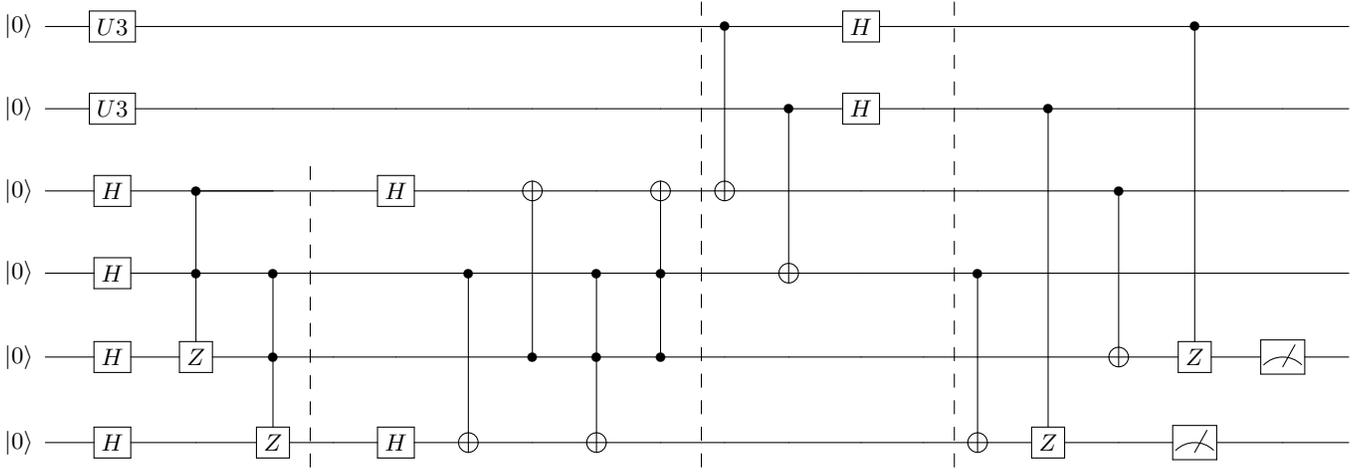
\begin{figure*}
    \centering
\Qcircuit @C=1.8em @R=2.1em {
&\lstick{\ket{0}} &\gate{U3}&\qw &\qw &\qw &\qw &\qw &\qw &\qw&\barrier{5}\qw&\ctrl{2}&\qw&\gate{H}&\barrier{5}\qw&\qw&\qw&\qw&\ctrl{4}&\qw&\qw \\
&\lstick{\ket{0}} &\gate{U3}&\qw &\qw &\qw &\qw &\qw &\qw &\qw &\qw&\qw&\ctrl{2}&\gate{H}&\qw&\qw&\ctrl{4}&\qw&\qw&\qw&\qw\\
&\lstick{\ket{0}} &\gate{H} & \ctrl{1}&\qw\barrier{3}\qw&\qw&\gate{H}&\qw&\targ&\qw&\targ&\targ&\qw&\qw&\qw&\qw&\qw&\ctrl{2}&\qw&\qw&\qw\\
&\lstick{\ket{0}} &\gate{H}&\ctrl{1}&\ctrl{1}&\qw&\qw&\ctrl{2}&\qw&\ctrl{1}&\ctrl{-1}&\qw&\targ&\qw&\qw&\ctrl{2}&\qw&\qw&\qw&\qw&\qw\\
& \lstick{\ket{0}} &\gate{H} &\gate{Z}&\ctrl{1} &\qw &\qw&\qw&\ctrl{-2}&\ctrl{1}&\ctrl{-1}&\qw&\qw&\qw&\qw&\qw&\qw&\targ&\gate{Z}&\meter&\qw\\
&\lstick{\ket{0}} &\gate{H} &\qw &\gate{Z} &\qw&\gate{H}&\targ&\qw&\targ&\qw&\qw&\qw&\qw&\qw&\targ&\gate{Z}&\qw&\meter&\qw&\qw
}
    \caption{A generalised circuit for quantum teleportation of arbitrary two-qubit state using a three-uniform-four-qubit hypergraph state.}
    \label{Fig4}
\end{figure*}
In this case, we will take a three-uniform four-qubit hypergraph state which we use as a quantum channel to teleport a two-qubit arbitrary state between Alice and Bob,
This four-qubit hypergraph state is shared between Alice and Bob's qubit, given as,
\begin{eqnarray}
\ket{h_4}=\frac{1}{4}(\ket{0000}+\ket{0001}+\ket{0010}+\ket{0011}\nonumber\\+\ket{0100}+\ket{0101}+\ket{0110}-\ket{0111}\nonumber\\+\ket{1000}+\ket{1001}+\ket{1010}+\ket{1011}\nonumber\\+\ket{1100}+\ket{1101}-\ket{1110}+\ket{1111})\nonumber\\
\label{RSP_Eq7}
\end{eqnarray}
After reducing this state, applying two Hadamard gate, two CNOT and two CCNOT gates as shown in Fig. \ref{Fig3} so the state become,
\begin{eqnarray}
\ket{H_4}_{a_1a_2b_1b_2}=\frac{1}{2}(\Ket{0000}+\ket{0101}+\ket{1010}+\ket{1111})\nonumber\\
\label{RSP_Eq8}
\end{eqnarray}
Where, the first two qubits $a_1$ and $a_2$ belong to Alice, and the last two qubits $b_1$ and $b_2$ belong to Bob, as denoted in the subscript.\\
Here in this protocol, we will teleport any two-qubit state         $\ket{\Psi}_{12}=(\alpha\ket{00}+\beta\ket{01}+\gamma\ket{10}+\delta\ket{11})$, where $|\alpha|^2+|\beta|^2+|\gamma|^2+|\delta|^2= 1$, Alice wants to teleport this arbitrary state. Initially, she has this unknown state and will combine it with the four-qubit hypergraph state of her share. So, the joint state of the arbitrary two-qubit state and four-qubit hypergraph state is written as,  $\ket{\Psi}_{12a_1a_2b_1b_2}=\ket{\Psi}_{12}\otimes\ket{\Psi}_{a_1a_2b_1b_2}$.

For this teleportation protocol \cite{kumar2020experimental}, Alice starts the process of the scheme by performing two CNOT operations on the combined state on qubits (1, $a_1$) and (2, $a_2$), where qubits ‘$1$’ and ‘$2$’ works as controlled qubits and qubits ‘$a_1$’ and ‘$a_2$’ works as target qubits. Further, Alice applies two Hadamard gates on qubits ‘$1$’ and ‘$2$’. Then she will measure her qubits on a computational basis and communicate the output of her outcome via a classical channel with Bob. After getting four-bit classical information from Alice, Bob can obtain the arbitrary two-qubit state at his location by performing appropriate unitary transformations on his qubits ‘$b_1$’ and ‘$b_2$’. The general quantum circuit for the teleportation of a two-qubit arbitrary state via a three-uniform four-qubit hypergraph state is shown in Fig. \ref{Fig4}.\\
\textbf{Circuit decomposition}: The four-qubit hypergraph state from Eq. \ref{RSP_Eq8} is given by $\ket{H_4}_{a_1a_2b_1b_2}$ and $\ket{\Psi}_{12}=(\alpha\ket{00}+\beta\ket{01}+\gamma\ket{10}+\delta\ket{11})$ gives the arbitrary two-qubit state which Alice wish to teleport. After CNOT gates on qubits ($1,a_1$) and ($1,a_2$) the joint state is changed to,
\begin{eqnarray}
\ket{\Psi'}_{12a_1a_2b_1b_2}=\frac{1}{2}[\alpha\ket{00}\otimes\ket{\chi_1}+\beta\ket{01}\otimes\ket{\chi_2}\nonumber\\+\gamma\ket{10}\otimes\ket{\chi_3}+\delta\ket{11}\otimes\ket{\chi_4}]\nonumber\\
\label{RSP_Eq9}
\end{eqnarray}
where,
\begin{eqnarray}
\ket{\chi_1}=(\ket{0000}+\ket{0101}+\ket{1010}+\ket{1111})_{a_1a_2b_1b_2}\nonumber\\
\ket{\chi_2}=(\ket{0100}+\ket{0001}+\ket{1110}+\ket{1011})_{a_1a_2b_1b_2}\nonumber\\
\ket{\chi_3}=(\ket{1000}+\ket{1101}+\ket{0010}+\ket{0111})_{a_1a_2b_1b_2}\nonumber\\
\ket{\chi_4}=(\ket{1100}+\ket{1001}+\ket{0110}+\ket{0011})_{a_1a_2b_1b_2}\nonumber\\
\label{RSP_Eq10}
\end{eqnarray}
Now, Alice applies Hadamard gates on qubits 1 and qubit 2 to covert the joint state into,
\begin{eqnarray}
\ket{\Psi''}_{12a_1a_2b_1b_2}=\frac{1}{2}[\alpha(\ket{00}+\ket{01}+\ket{10}+\ket{11})\otimes\ket{\chi_1}\nonumber\\+\beta(\ket{00}-\ket{01}+\ket{10}-\ket{11})\otimes\ket{\chi_2}\nonumber\\+\gamma(\ket{00}+\ket{01}-\ket{10}-\ket{11})\otimes\ket{\chi_3}\nonumber\\+\delta(\ket{00}-\ket{01}-\ket{10}+\ket{11})\otimes\ket{\chi_4}]\nonumber\\
\label{RSP_Eq11}
\end{eqnarray}
\begin{table*}
\begin{tabular}{|c|c|c|} 
\hline
Alice's outcome & Bob obtains & Bob's Operations\\
\hline
$\Ket{0000}$ & $(\alpha\ket{00}+\beta\ket{01}+\gamma\ket{10}+\delta\ket{11})$ & $I_{b_1}I_{b_2}$\\
$\Ket{0001}$ & $(\alpha\ket{01}+\beta\ket{00}+\gamma\ket{11}+\delta\ket{10})$ & $I_{b_1}X_{b_2}$\\
$\Ket{0010}$ & $(\alpha\ket{10}+\beta\ket{11}+\gamma\ket{00}+\delta\ket{01})$ & $X_{b_1}I_{b_2}$\\
$\Ket{0011}$ & $(\alpha\ket{11}+\beta\ket{10}+\gamma\ket{01}+\delta\ket{00})$ & $X_{b_1}X_{b_2}$\\
$\Ket{0100}$ & $(\alpha\ket{00}-\beta\ket{01}+\gamma\ket{10}-\delta\ket{11})$ & $I_{b_1}Z_{b_2}$\\
$\Ket{0101}$ & $(\alpha\ket{01}-\beta\ket{00}+\gamma\ket{11}-\delta\ket{10})$ & $I_{b_1}ZX_{b_2}$\\
$\Ket{0110}$ & $(\alpha\ket{10}-\beta\ket{11}+\gamma\ket{00}-\delta\ket{01})$ & $X_{b_1}Z_{b_2}$\\
$\Ket{0111}$ & $(\alpha\ket{11}-\beta\ket{10}+\gamma\ket{01}-\delta\ket{00})$ & $X_{b_1}ZX_{b_2}$\\
$\Ket{1000}$ & $(\alpha\ket{00}+\beta\ket{01}-\gamma\ket{10}-\delta\ket{11})$ & $Z_{b_1}I_{b_2}$\\
$\Ket{1001}$ & $(\alpha\ket{01}+\beta\ket{00}-\gamma\ket{11}-\delta\ket{10})$ & $Z_{b_1}X_{b_2}$\\
$\Ket{1010}$ & $(\alpha\ket{10}+\beta\ket{11}-\gamma\ket{00}-\delta\ket{01})$ & $ZX_{b_1}I_{b_2}$\\
$\Ket{1011}$ & $(\alpha\ket{11}+\beta\ket{10}-\gamma\ket{01}-\delta\ket{00})$ & $ZX_{b_1}X_{b_2}$\\
$\Ket{1100}$ & $(\alpha\ket{00}-\beta\ket{01}-\gamma\ket{10}+\delta\ket{11})$ & $Z_{b_1}Z_{b_2}$\\
$\Ket{1101}$ & $(\alpha\ket{01}-\beta\ket{00}-\gamma\ket{11}+\delta\ket{10})$ & $Z_{b_1}ZX_{b_2}$\\
$\Ket{1110}$ & $(\alpha\ket{10}-\beta\ket{11}-\gamma\ket{00}+\delta\ket{01})$ & $ZX_{b_1}Z_{b_2}$\\
$\Ket{1111}$ & $(\alpha\ket{11}-\beta\ket{10}-\gamma\ket{01}+\delta\ket{00})$ & $ZX_{b_1}ZX_{b_2}$\\
\hline
\end{tabular}
\caption{Bob’s unitary operations corresponding to Alice’s measurement results. For example, let Alice measures her qubit, and it turns out to be $
\ket{0100}$ then she tells Bob about her qubit classically, when Bob gets it, he applies identity and Pauli Z operation on his qubits $b_1$ and $b_2$, respectively. If he gets, say $\ket{1011}$, he applies Pauli XZ (XZ = iY) and X operations on his qubits $b_1$ and $b_2$. In this way, Bob can recover the actual arbitrary state at his location. So, this teleportation scheme is deterministic, and the success probability is 100\%.}
\label{TableII}
\end{table*}
After expanding and rearranging the above equation, we get\\
\begin{eqnarray}
\ket{\Psi''}=\frac{1}{4}[\ket{0000}\otimes\ket{\Psi^1}+\ket{0001}\otimes\ket{\Psi^2}+\ket{0010}\otimes\ket{\Psi^3}\nonumber\\+\ket{0011}\otimes\ket{\Psi^4}+\ket{0100}\otimes\ket{\Psi^5}+\ket{0101}\otimes\ket{\Psi^6}\nonumber\\+\ket{0110}\otimes\ket{\Psi^7}+\ket{0111}\otimes\ket{\Psi^8}+\ket{1000}\otimes\ket{\Psi^9}\nonumber\\+\ket{1001}\otimes\ket{\Psi^{10}}+\ket{1010}\otimes\ket{\Psi^{11}}+\ket{1011}\otimes\ket{\Psi^{12}}\nonumber\\+\ket{1100}\otimes\ket{\Psi^{13}}+\ket{1101}\otimes\ket{\Psi^{14}}+\ket{1110}\otimes\ket{\Psi^{15}}\nonumber\\+\ket{1111}\otimes\ket{\Psi^{16}}]_{12a_1a_2b_1b_2}\nonumber\\
\label{RSP_Eq12}
\end{eqnarray}
where,
\begin{eqnarray}
\ket{\Psi^1}=(\alpha\ket{00}+\beta\ket{01}+\gamma\ket{10}+\delta\ket{11})_{b_1b_2}\nonumber\\
\ket{\Psi^2}=(\alpha\ket{01}+\beta\ket{00}+\gamma\ket{11}+\delta\ket{10})_{b_1b_2}\nonumber\\
\ket{\Psi^3}=(\alpha\ket{10}+\beta\ket{11}+\gamma\ket{00}+\delta\ket{01})_{b_1b_2}\nonumber\\
\ket{\Psi^4}=(\alpha\ket{11}+\beta\ket{10}+\gamma\ket{01}+\delta\ket{00})_{b_1b_2}\nonumber\\
\ket{\Psi^5}=(\alpha\ket{00}-\beta\ket{01}+\gamma\ket{10}-\delta\ket{11})_{b_1b_2}\nonumber\\
\ket{\Psi^6}=(\alpha\ket{01}-\beta\ket{00}+\gamma\ket{11}-\delta\ket{10})_{b_1b_2}\nonumber\\
\ket{\Psi^7}=(\alpha\ket{10}-\beta\ket{11}+\gamma\ket{00}-\delta\ket{01})_{b_1b_2}\nonumber\\
\ket{\Psi^8}=(\alpha\ket{11}-\beta\ket{10}+\gamma\ket{01}-\delta\ket{00})_{b_1b_2}\nonumber\\
\ket{\Psi^9}=(\alpha\ket{00}+\beta\ket{01}-\gamma\ket{10}-\delta\ket{11})_{b_1b_2}\nonumber\\
\ket{\Psi^{10}}=(\alpha\ket{01}+\beta\ket{00}-\gamma\ket{11}-\delta\ket{10})_{b_1b_2}\nonumber\\
\ket{\Psi^{11}}=(\alpha\ket{10}+\beta\ket{11}-\gamma\ket{00}-\delta\ket{01})_{b_1b_2}\nonumber\\
\ket{\Psi^{12}}=(\alpha\ket{11}+\beta\ket{10}-\gamma\ket{01}-\delta\ket{00})_{b_1b_2}\nonumber\\
\ket{\Psi^{13}}=(\alpha\ket{00}-\beta\ket{01}-\gamma\ket{10}+\delta\ket{11})_{b_1b_2}\nonumber\\
\ket{\Psi^{14}}=(\alpha\ket{01}-\beta\ket{00}-\gamma\ket{11}+\delta\ket{10})_{b_1b_2}\nonumber\\
\ket{\Psi^{15}}=(\alpha\ket{10}-\beta\ket{11}-\gamma\ket{00}+\delta\ket{01})_{b_1b_2}\nonumber\\
\ket{\Psi^{16}}=(\alpha\ket{11}-\beta\ket{10}-\gamma\ket{01}+\delta\ket{00})_{b_1b_2}\nonumber\\
\label{RSP_Eq13}
\end{eqnarray}
Now Alice measures her qubit on a computational basis and sends the four-bit classical information to Bob. Depending upon Alice’s classical bits, Bob performs unitary transformations on his qubits to obtain the actual arbitrary two-qubit state at his location, as shown in Table \ref{TableII}.
\section{Experimental realization of QT in IBM QE\label{RSP_Sec4}}
\begin{figure*}
    \centering
    \includegraphics[width=1\textwidth]{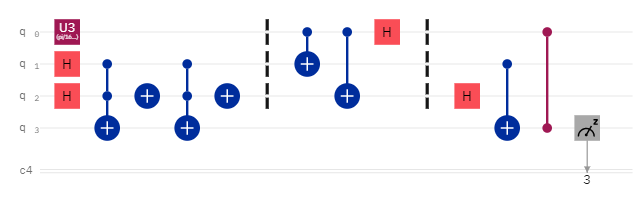}
    \caption{Quantum circuit to perform quantum teleportation of single-qubit state using three-qubit hypergraph state implemented in the IBM quantum computer with minimum number of gates.}
    \label{Fig5}
\end{figure*}

‘IBM Quantum Experience (QE)’\cite{walter2005} is the platform where we can perform quantum teleportation. The actual circuits for single-qubit and two-qubit teleportation are given in Fig. \ref{Fig5} and Fig. \ref{Fig8}, respectively, in the IBM Quantum Experience processor. We have chosen the required gates from the toolbox, where we have tried to minimize the no. of gates required to implement a hypergraph state to reduce the gate errors in our experiment.
To run the quantum circuit in IBM QE, we can choose different numbers of shots like 2048, 4096, 8192 etc., where shot means the number of times the experiment is executed in the quantum processor. If we choose a significantly less no. of shots (like 5, 10, 15), we will not get any favourable result compared to a large no. of shots. In IBM QE, the maximum no. of shots with which we can run is 8192; this gives an almost accurate result of an experiment.\\
\section{Quantum state tomography\label{RSP_Sec5}}
Quantum state tomography\citep{JDFV2001,Gotteman2000,altepeter2004,niggebaum2011} is a well-known approach to characterise the quantum state by comparing collections of the theoretical and experimental density matrices. We can obtain the accuracy of our teleportation protocol by comparing theoretical and experimental density matrices. The theoretical density matrix of the quantum given arbitrary state can be written as,
\begin{eqnarray}
\rho^T=\ket{\Psi}\bra{\Psi}
\label{RSP_Eq14}
\end{eqnarray}
The experimental density matrix\cite{SMST2017,Childs2001,Vishnu2018} for a multi-qubit state is given by,
\begin{eqnarray}
\rho^E=\frac{1}{2^N}\sum_{k_1,k_2,k_3..k_N=0}^{3}T_{k_1,k_2,k_3..k_N}(\sigma_{k_1}\otimes\sigma_{k_2}\nonumber\\\otimes\sigma_{k_3}..\otimes\sigma_{k_N})\nonumber\\
\label{RSP_Eq15}
\end{eqnarray}
where, $\sigma_{k_i}$'s are the the Pauli matrices $i\in \{1,2,3...,N\}$ and $T_{k_1,k_2,k_3...k_N}$ represents the particular projective measurement in the experiment given as, 
\begin{eqnarray}
T_{k_1,k_2,k_3...k_N}=S_{k_1}\times S_{k_2}\times S_{k_3}...\times S_{k_N}
\label{RSP_Eq16}
\end{eqnarray}
Where $S_{k_i}$ represents the Stocks parameter\cite{Vishnu2018} in the Bloch sphere,$k_1$,$k_2$,$k_3$...$k_N$ can have values 0, 1, 2 and 3, representing quantum gates I, X, Y, Z, respectively.
\section{Result\label{RSP_Sec6}}
In this section, we will demonstrate our results for both the case, namely an arbitrary single-qubit state using a three-uniform three-qubit hypergraph state and an arbitrary two-qubit state using a three-uniform four-qubit hypergraph state. We first verify our result in the ``ibmq\_qasm\_simulator" (with 8192 shots to make our result more accurate and to reduce the statistical errors). Further, we run the result in ``ibmq\_16\_melbourne” and compare both the results; we find that ``ibmq\_16\_melbourne” has occurred some errors due to the decoherence\cite{Harper} effect in the quantum channel. Finally, we have performed the teleportation scheme for both single and two-qubit states in ``ibmq\_qasm\_simulator" and ``ibmq\_16\_melbourne” (real device), and the probability obtained is shown in Fig. \ref{Fig6} and Fig. \ref{Fig9}.\\
\textbf{Note}: The circuits (Fig. \ref{Fig5} and Fig. \ref{Fig8}) shown here are drawn on the IBM circuit drawer; these are similar to all the operations discussed in the text and the circuit shown in Fig. \ref{Fig2} and Fig. \ref{Fig4}. Here confusion might arise between the “circuits operations” and “Bob’s operations”, as shown in Table \ref{TableI} and Table \ref{TableII}. However, as seen from quantum circuits in Figs. \ref{Fig5} and \ref{Fig8}, it is shown that CZ and CX operations are applied between Alice qubit and Bob’s qubits. These operations are performed locally, but both Alice and Bob are situated far apart. We have operated it just for convenience to get the appropriate results.
\subsection{Result for teleportation of single-qubit state}
\begin{figure*}
    \centering
    \includegraphics[width=0.5\textwidth]{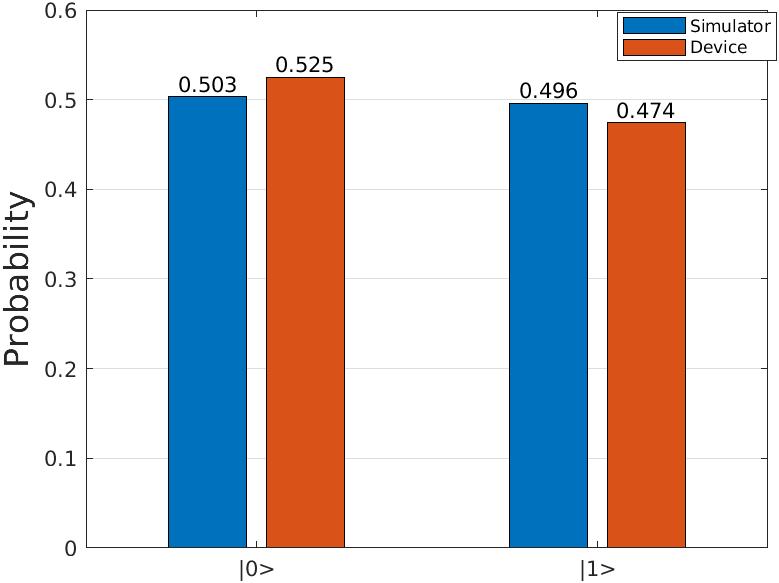}
    \caption{Histogram. This histogram shows the comparison between the probabilities obtained by ``ibmq\_qasm\_simulator" and ``ibmq\_16\_melbourne” for teleportation of a single-qubit state.}
    \label{Fig6}
\end{figure*}

Let us take a single-qubit state to analyse our experimental result $\ket{\Psi}_a=\frac{1}{\sqrt{2}}(\ket{0}+\ket{1})$ this qubit can be made by setting the parameter of the $U_3(\pi/2,0,0)$ gate; we first run our circuit on the ``ibmq\_qasm\_simulator" then on ``ibmq\_16\_melbourne” with 8192 shot each. A comparison of the result by taking the average probability of 5 runs for each has been shown in Fig. \ref{Fig6}. Then, we will calculate the fidelity of the circuit to know how well the state is teleported. The following formula gives the fidelity\cite{nielsen2002quantum,Pathak2019},
\begin{eqnarray}
F(\rho^T, \rho^E)= Tr(\sqrt{\sqrt{\rho^T}\rho^E\sqrt{\rho^T}})
\label{RSP_Eq17}
\end{eqnarray}
\begin{figure*}
\begin{subfigure}
    \centering
    \includegraphics[width=.4\linewidth]{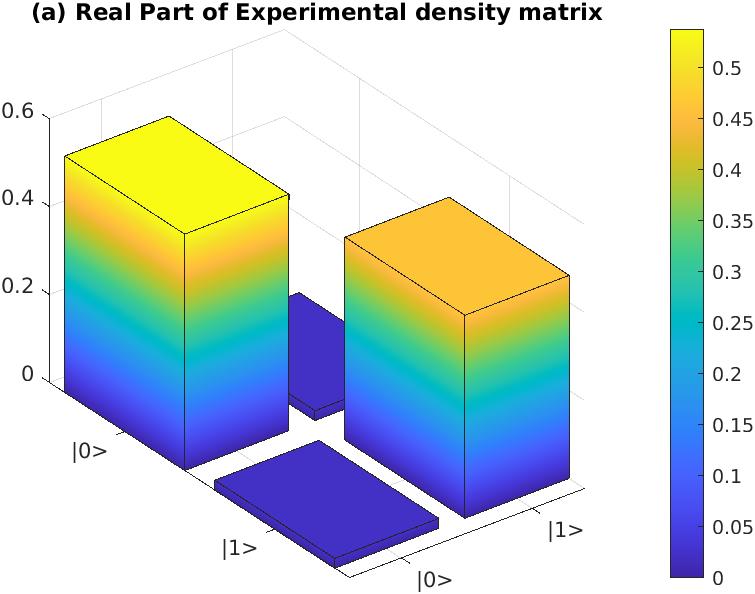}
\end{subfigure}
\begin{subfigure}
    \centering
    \includegraphics[width=.4\linewidth]{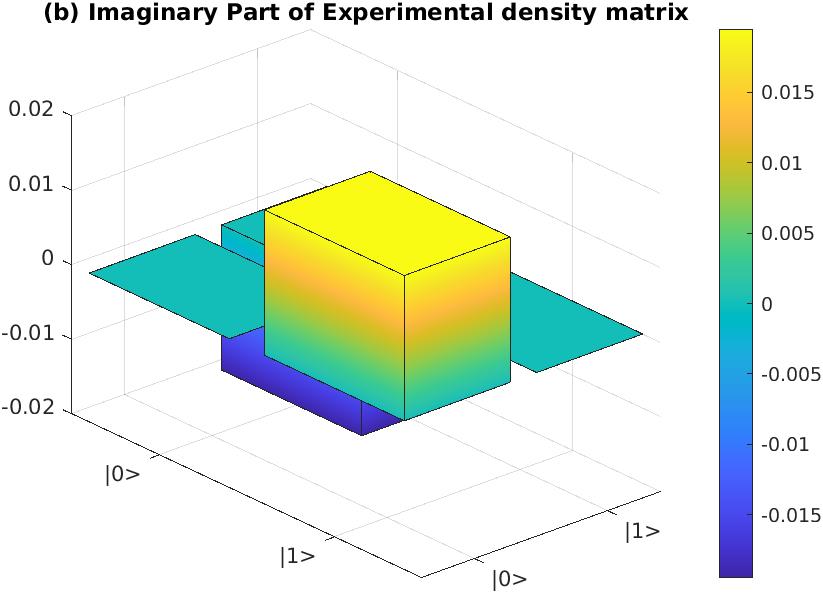}
\end{subfigure}
\newline
\begin{subfigure}
    \centering
    \includegraphics[width=.4\linewidth]{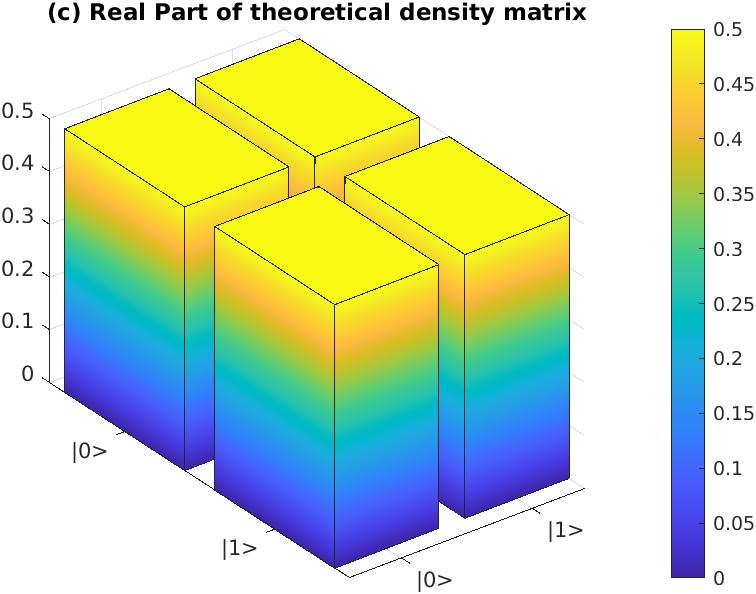}
\end{subfigure}
\begin{subfigure}
    \centering
    \includegraphics[width=.4\linewidth]{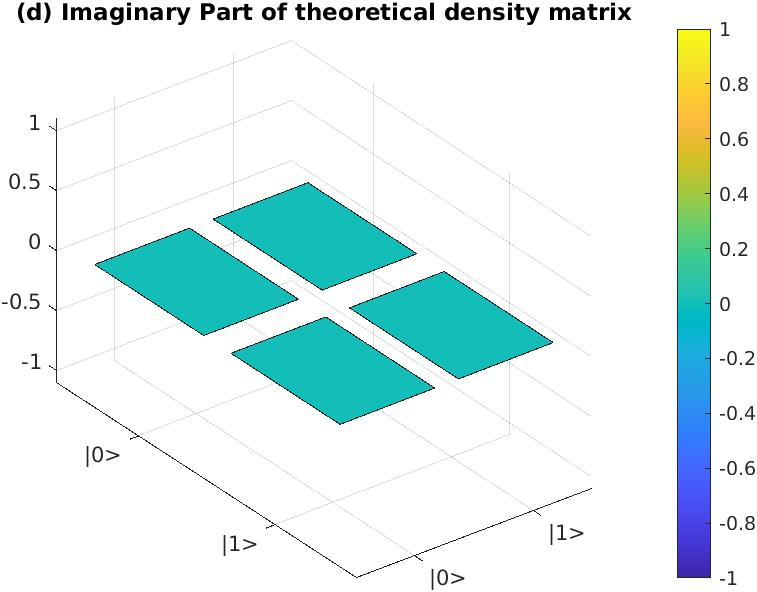}
\end{subfigure}
    \caption{Real and Imaginary part of Experimental and Theoretical density matrix for teleportation of single-qubit state $\ket{\Psi}=\frac{1}{\sqrt{2}}(\ket{0})+\ket{1})$, (a) Real part of the experimental density matrix, (b) Imaginary part of experimental density matrix, (c) Real part of theoretical density matrix, and (d) Imaginary part of the theoretical density matrix. These results are obtained from the ``ibmq\_16\_melbourne” device.}
    \label{Fig7}
\end{figure*}
\begin{figure*}
    \centering
    \includegraphics[width=1\linewidth]{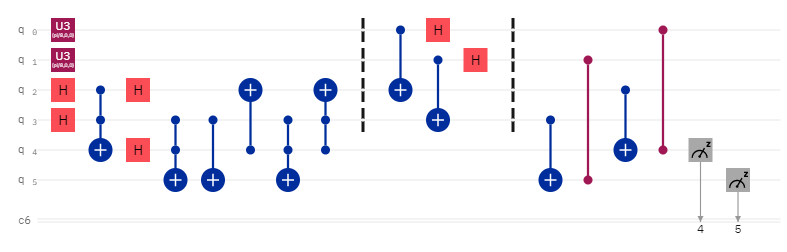}
    \caption{Quantum circuit to perform quantum teleportation of two-qubit state using three-uniform four-qubit hypergraph state implemented in the IBM quantum computer.}
    \label{Fig8}
\end{figure*}
\begin{figure*}
    \centering
    \includegraphics[width=.5\linewidth]{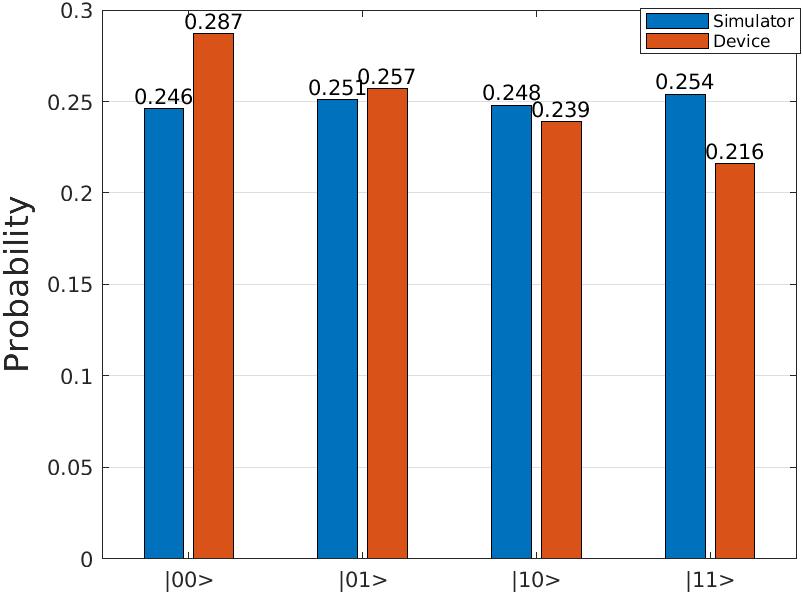}
    \caption{Histogram. This histogram shows the comparison between the probabilities obtained by ``ibmq\_qasm\_simulator" and ``ibmq\_16\_melbourne” for teleportation of a two-qubit state.}
    \label{Fig9}
\end{figure*}
here, $\rho^T$ and $\rho^E$ reperesents the theoretical and experimental density matrices respectively.\\
Theoretical density matrix for the given single-qubit state is obtained as,
\begin{eqnarray}
\rho^T=\frac{1}{2}\begin{pmatrix}
 1    & 1 \\
  1   & 1
\end{pmatrix}
\label{RSP_Eq18}
\end{eqnarray}
and experimental density matrix can be written according to Eq. \ref{RSP_Eq15}
\begin{eqnarray}
\rho^E=\frac{1}{2}[T_I(I)+T_X(\sigma_X)+T_Y(\sigma_Y)+T_Z(\sigma_Z)]\nonumber\\ \label{RSP_Eq19}
\end{eqnarray}
Where, $\sigma_X$, $\sigma_Y$, $\sigma_Z$ are Pauli matrices and I is the identity matrix. $T_{k_i}=S_{k_i}$ where, stokes parameters are given by $S_I= P_{\ket{0I}}+P_{\ket{1I}}$, $S_X=P_{\ket{0X}}-P_{\ket{1X}}$, $S_Y= P_{\ket{0Y}}- P_{\ket{1Y}}$ and $S_Z= P_{\ket{0Z}}- P_{\ket{1Z}}$. Here $P_{\ket{0k_i}}$ represents the probability of the qubit to be found in state $\ket{0}$, and it is measured in $k_i$ basis, and $P_{\ket{1k_i}}$ represents the probability of the qubit to be found in state $\ket{1}$, and it is measured in $k_i$ basis.\\
So, the experimental density matrix can be written for the given single-qubit state as,
\begin{eqnarray}
\rho^E=\begin{pmatrix}
 0.5380    & 0.0225 \\
  0.0225   & 0.4620
\end{pmatrix}+i\begin{pmatrix}
 0    & -0.0195 \\
  0.0195   & 0
\end{pmatrix}
\nonumber\\
\label{RSP_Eq20}
\end{eqnarray}
We have calculated the fidelity between the theoretical density matrix (\ref{RSP_Eq16}) and the experimental density matrix (\ref{RSP_Eq18}) to be F($\rho^T$, $\rho^E$) = 0.7228
\subsection{Result for teleportation of two-qubit state}
\begin{figure*}
\begin{subfigure}
    \centering
    \includegraphics[width=.45\linewidth]{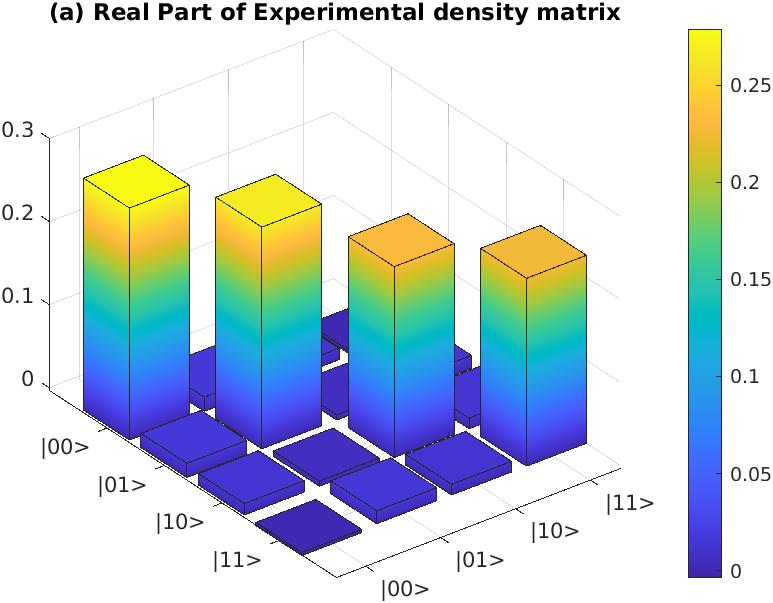}
\end{subfigure}
\begin{subfigure}
    \centering
    \includegraphics[width=.45\linewidth]{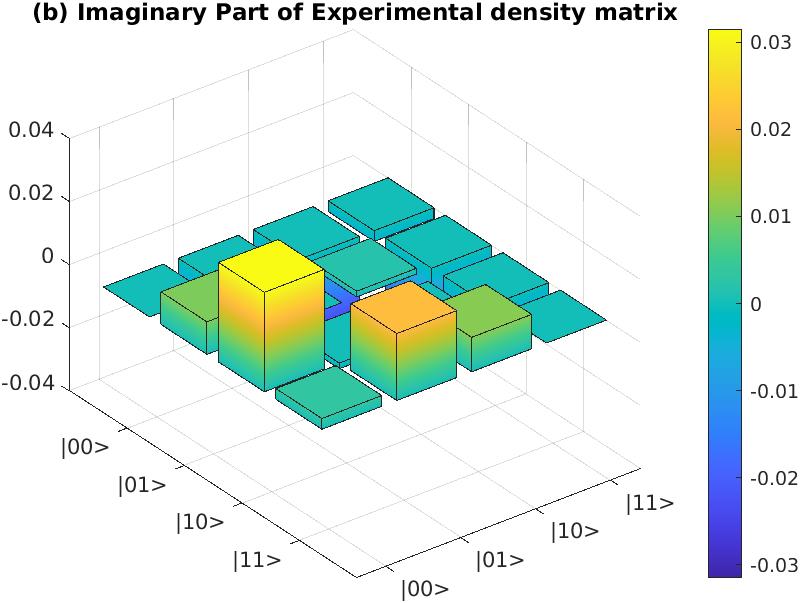}
\end{subfigure}
\newline
\begin{subfigure}
    \centering
    \includegraphics[width=.45\linewidth]{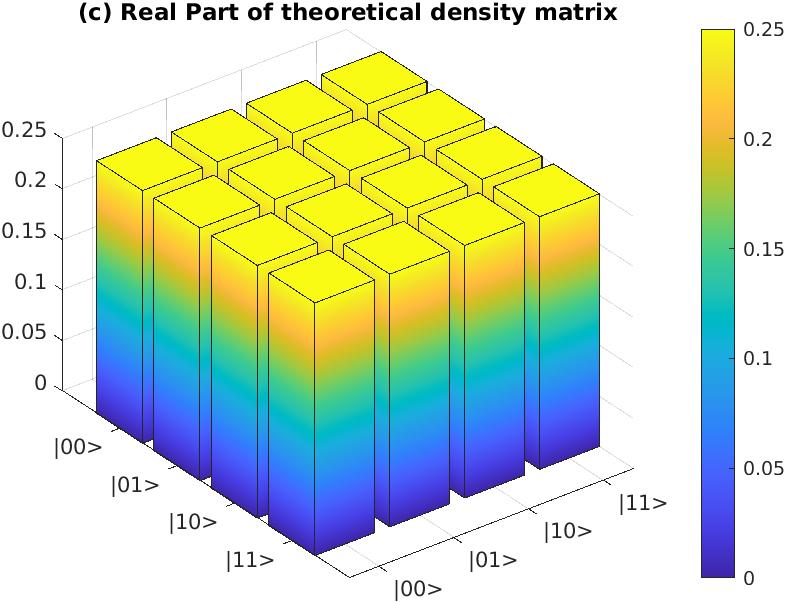}
\end{subfigure}
\begin{subfigure}
    \centering
    \includegraphics[width=.45\linewidth]{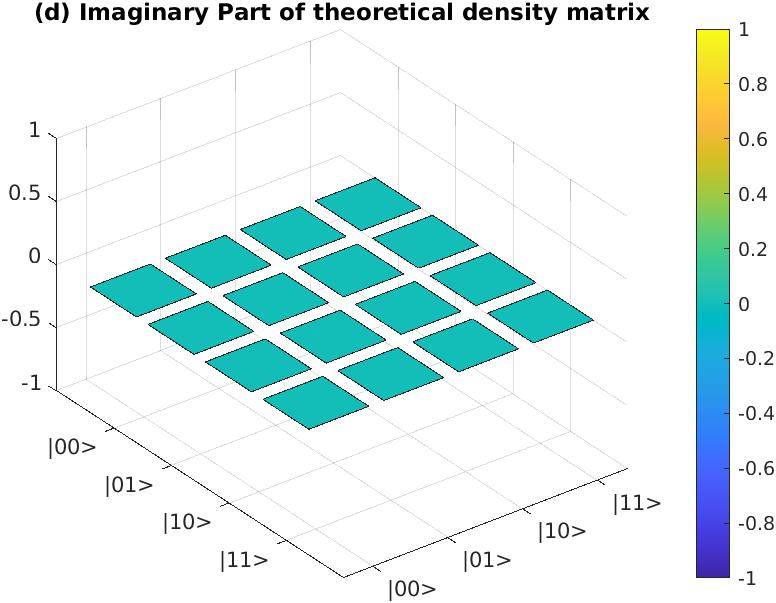}
\end{subfigure}
    \caption{Real and Imaginary part of Experimental and Theoretical density matrix for teleportation of two-qubit state $\ket{\Psi}=\frac{1}{2}(\ket{00})+\ket{01}+\ket{10}++\ket{11})$ , (a) Real part of the experimental density matrix, (b) Imaginary part of the experimental density matrix, (c) Real part of the theoretical density matrix, (d) Imaginary part of the theoretical density matrix. These results are obtained from the ``ibmq\_16\_melbourne” device.}
    \label{Fig10}
\end{figure*}
We perform the same procedure as discussed for single qubit-teleportation, i.e., we first run the quantum circuit on ``ibmq\_qasm\_simulator” and then on ``ibmq\_16\_melbourne” each for 8192 shots for more accuracy. The comparison of the probability obtained is shown in Fig. \ref{Fig9}; for ``ibmq\_qasm\_simulator”, the probability of getting each state is almost the same; it performs nearly to the theoretical result. Whereas, in the case of ``ibmq\_16\_melbourne” (real device), it shows that the probability of getting each possible state for the two-qubit state is different here errors present due to noises in the quantum channel. Noise appears due to decoherence in the quantum channel, gate errors and state preparation errors.\\
Let us take a two-qubit state given by $\Ket{\Psi}_{12}=\frac{1}{2}(\ket{00}+\ket{01}+\ket{10}+\ket{11})$.\\
The theoretical density matrix for this state can be written as,
\begin{eqnarray}
\rho^E=\frac{1}{4}\begin{pmatrix}
 1    & 1 & 1 & 1 \\
 1    & 1 & 1 & 1 \\
  1    & 1 & 1 & 1 \\
 1    & 1 & 1 & 1 
\end{pmatrix}
\label{RSP_Eq21}
\end{eqnarray}
Besides, the experimental density matrix for teleportation of a given two-qubit state\cite{sisodia2017teleportation} can be written according to Eq. \ref{RSP_Eq13} as,
\begin{eqnarray}
\rho^E=\frac{1}{4}[T_{II}(I\otimes I)+T_{IX}(I\otimes\sigma_X)+T_{XI}(\sigma_X\otimes I)\nonumber\\+T_{IY}(I\otimes\sigma_Y)+T_{YI}(\sigma_Y\otimes I)+T_{IZ}(I\otimes\sigma_Z)\nonumber\\+T_{ZI}(\sigma_Z\otimes I)+T_{XY}(\sigma_X\otimes\sigma_Y)+T_{YX}(\sigma_Y\otimes\sigma_X)\nonumber\\+T_{YZ}(\sigma_Y\otimes\sigma_Z)+T_{ZY}(\sigma_Z\otimes\sigma_Y)+T_{XZ}(\sigma_X\otimes\sigma_Z)\nonumber\\+T_{ZX}(\sigma_Z\otimes\sigma_X)+T_{XX}(\sigma_X\otimes\sigma_X)+T_{YY}(\sigma_Y\otimes\sigma_Y)\nonumber\\+T_{ZZ}(\sigma_Z\otimes\sigma_Z)]\nonumber\\
\label{RSP_Eq22}
\end{eqnarray}
where, $\sigma_{k_i}$'s are Pauli matrices and $T_{k_1k_2}=S_{k_1}\times S_{k_2}$, here, $S_{k_i}$ is Stokes parameter, as discussed earlier.\\
The experimental density matrix is calculated after running the circuit in “IBM 16 Melbourne” and after measurement on different bases and finding different probability distribution of all possible states.\\
So, the experimental density matrix for the given two-qubit state is given by,
\begin{eqnarray}
\rho^E=\begin{pmatrix}
   0.2790  & 0.0170 & 0.0136 & -0.0038 \\
    0.0170  & 0.2666 & 0.0063 & 0.0155 \\
  0.0136   & 0.0063 & 0.2290 & 0.0128 \\
  -0.0038 & 0.0155 & 0.0128 & 0.2254
\end{pmatrix}\nonumber\\+i\begin{pmatrix}
 0    & -0.0103 & -0.0315 & -0.0034 \\
  0.0103   & 0 & 0.0018 & -0.0212 \\
  0.0315 & -0.0018 & 0 & -0.0109 \\
  0.0034 & 0.0212 & 0.0109 & 0
\end{pmatrix}
\label{RSP_Eq23}
\end{eqnarray}
Thereby, the fidelity between the theoretical and the experimental density matrix is calculated to be, 0.5298.
\section{Conclusion\label{RSP_Sec7}}
To conclude, here we have implemented quantum circuits in the IBM quantum computer and demonstrated the teleportation of single-qubit and two-qubit arbitrary states using three-uniform, three-qubit and four-qubit ``hypergraph states”, respectively. We here analysed theoretical methods for the teleportation of arbitrary states. The appropriate unitary operations required at Bob’s end to make teleportation successful is shown for single and two-qubit states in Table \ref{TableI} and Table \ref{TableII}. We have successfully run the above quantum circuits in the ``ibmq\_qasm\_simulator” and ``ibmq\_16\_melbourne”(real device) and compared the results using density matrices; the comparison of probability obtained in the real device to the simulator is shown in the histogram plots. The quantum simulator performs almost as theoretical results without any error, whereas the real device shows faults due to decoherence effect, state preparation error, gate errors, readout error, etc.

\end{document}